\documentclass[preprint]{aastex}
\usepackage{epsfig}
\usepackage{enumerate}
\def\beq{\begin{equation}}
\def\eeqno#1{\label{#1}\end{equation}}

\def\rarrow{\rightarrow }

\def\dleft{\rlap{{\it D}}\raise 8pt
\hbox{$\scriptscriptstyle\Leftarrow$}}
\def\dright{\rlap{{\it
D}}\raise 8pt\hbox{$\scriptscriptstyle\Rightarrow$}}

\def\kms{{\rm km~s^{-1}}}
\def\cmss{{\rm cm~s^{-2}}}
\def\kpc{~{\rm Kpc}}
\def\mpc{~{\rm Mpc}}

\def\msun{M_{\odot}}
\def\az{a_{0}}

\def\lsun{L_{\odot}}
\def\l0{\ell_{0}}

\def\rar{\rightarrow}

\def\l{\lambda}

\def\m{\mu}

\def\xlimin{{x\rarrow\infty \atop{\raise 1pt\hbox to 30pt
{\rightarrowfill}}}}
\def\limlim#1#2{{#1\rarrow #2 \atop{\raise 1pt\hbox to 30pt
{\rightarrowfill}}}}

\def\vg{{\bf g}}

\def\gN{g\_N}

\def\m{\mu}

\def\_#1{_{\scriptscriptstyle #1}}
\def\^#1{^{\scriptscriptstyle #1}}

\begin{document}
\title{Ultra-diffuse cluster galaxies as key to the MOND cluster conundrum}
\author{Mordehai Milgrom}
\affil{Department of Particle Physics and Astrophysics, Weizmann Institute}

\begin{abstract}
MOND reduces greatly the mass discrepancy in clusters of galaxies, but does leave a global discrepancy of about a factor of 2 (epitomized by the structure of the Bullet Cluster). It has been proposed, within the minimalist and purist MOND, that clusters harbor some indigenous, yet undetected, cluster baryonic (dark) matter (CBDM), whose total amount is comparable with that of the observed hot gas. Koda et al. have recently identified more than a thousand ultra-diffuse, galaxy-like objects (UDGs) in the Coma cluster. These, they argue, require, within Newtonian dynamics, that they are much more massive than their observed stellar component. Here, I propound that some of the CBDM is internal to UDGs, which endows them with robustness. The rest of the CBDM objects formed in now-disrupted kin of the UDGs, and is dispersed in the intracluster medium. The discovery of cluster UDGs is not in itself a resolution of the MOND cluster conundrum, but it lends greater plausibility to CBDM as its resolution. Alternatively, if the UDGs are only now falling into Coma, their large size and very low surface brightness could result from the inflation due to the MOND, variable external-field effect (EFE). I also consider briefly solutions to the conundrum that invoke more elaborate extensions of purist MOND, e.g., that in clusters, the MOND constant takes up larger than canonical values of the MOND constant. Whatever solves the cluster conundrum within MOND might also naturally account for UDGs.
\end{abstract}
\maketitle
\section{Introduction}
MOND is a paradigm of dynamics that strives to explain away the mass discrepancies in galactic systems, and the Universe at large, without dark matter (DM). It departs from standard (e.g., Newtonian) dynamics at accelerations that are lower than the MOND acceleration constant $\az\approx 1.2\times 10^{-8}\cmss$. Recent reviews of MOND are Famaey \& McGaugh (2012) and Milgrom (2014).
\par
It has been known from the early 1990s that MOND, in its minimalist and purist form (which assumes, e.g., a universal value of $\az$), does not completely remove the mass discrepancies in galaxy clusters (e.g., The \& White 1988; Gerbal \& al. 1992; Sanders 1994; Sanders 1999; Sanders 2003; Pointecouteau \& Silk 2005; Takahashi \& Chiba 2007; Angus, Famaey, \& Buote 2008).
\par
The general situation is described in Milgrom (2008). It is succinctly this: Taking a cluster at large, say within $\sim 1-2\mpc$ of its center, MOND does reduce greatly the mass discrepancy compared with what is obtained with Newtonian dynamics. However, there remains a mass discrepancy of a factor of about 2. Namely, even with MOND we need clusters to contain some yet undetected component whose total mass is roughly the same as that of the presently known baryons in the clusters. This latter is mainly in the form of x-ray emitting hot gas, with stars in cluster galaxies contributing typically $\sim 10-20\%$ of that. Analyzing the radius dependence of the discrepancy (see the above references), one finds that this extra matter has to be distributed in a more centrally concentrated manner than the hot gas, rather like the galaxies. In other words, the remaining MOND discrepancies are more pronounced in the core, where they can reach a factor of ten or even more in some cases. Hypothetical, sterile neutrinos have been suggested as the constituents of this extra matter, but this is difficult to accommodate (Angus, Famaey, \& Buote 2008).
\par
There are other suggestions to avoid the remaining discrepancies within more elaborate schemes that embody MOND. For example, it has been suggested that $\az$ might not be universal, and that it is larger in cluster than it is in galaxies (e.g., Zhao \& Famaey 2012, Khoury 2015).
\par
Yet other suggestions are that MOND is somehow the emergent dynamics in galaxies induced by an omnipresent medium with specially designed properties, with enough flexibility to reproduce galactic MOND dynamics on one hand, but to have nonbaryonic-DM effects in galaxy clusters (e.g., Zhao and Li 2010; Ho, Minic, \& Ng, 2012; Berezhiani \& Khoury 2015). This hypothetical `DM' is anything but similar to the standard particle DM, which dominates the mainstream thinking.
\par
All such proposals depart in one way or another from the purist MOND.
\par
My own favorite solution retains purist MOND as modified dynamics with a (presently) universal value of $\az$, positing some yet undetected cluster {\it baryonic} `DM' (CBDM; Milgrom 2008). This is based on a. minimalist penchant, b. a feeling that we do not yet know everything about the baryonic contents of galaxy clusters, so CBDM is not too tall an order, and c. my past experience with MOND and clusters (see Sec. \ref{cbdm}).
\par
The known baryons in clusters make up only about $4\%$ of the baryon budget in the Universe (e.g., Fukugita \& Peebles 2004). Thus, the CBDM has to make a similar contribution to the baryon budget. In light of the fact that the whereabouts of $\gtrsim 50$ percent of the baryons in the Universe today are not known (the so-called `missing baryons'), it follows that only a small fraction of the still-missing baryons could account for all the needed CBDM.
\par
According to MOND, this CBDM is indigenous to clusters; so, it is not subject to the constraints we have on baryonic
DM candidates which assume that it makes all the DM in the Universe.
\par
It has also been suggested (Milgrom 2008) that through its interaction with the cluster gas, the CBDM could account for the heating of the cluster cores, alleviating, or even solving, the so-called cooling flow problem.
\par
The colliding `Bullet Cluster' is often adduced as strong evidence for DM, with the understood implication that it is evidence for {\it the} non-standard-model particle DM that is invoked by the mainstream to fill the Universe.
In the case of the Bullet, most of the baryons (in the form of hot gas) are found in the collision zone, while the `DM', as deduced from weak lansing, is mainly in two regions flanking the collision zone, where the galaxies from the colliding clusters are found, after they went through, hardly affected by the collision.
The claim then goes, that modified-dynamics alternatives to DM perforce predict that the `phantom DM' should appear where most of the baryons are, unlike what is seen in the Bullet
. This statement, in itself, is not correct. Modified dynamics theories, including MOND, do not predict that the discrepancies should follow the baryons.
It is true, though, that purist MOND does not account for the observed geometry of the Bullet without invoking
some yet undetected matter in the system. However, just as everything that glitters is not gold, everything that is dark is not {\it the} DM. What is `seen' in the Bullet might well be just an inkling of small amounts of yet undetected baryons indigenous to clusters.

Indeed, it turns out that the presence of CBDM as required in single clusters, automatically predicts the behavior seen in the Bullet.

So, have we learnt anything new about MOND from the Bullet? Its appearance tells us that whatever makes the cluster DM has to be able to cross the gas in the cluster core at least once without losing much momentum (as the galaxies do). However, this is already a known requirement from the CBDM deduced from single clusters. The crossing time in clusters is rather shorter than the Hubble time; so, to have survived in single clusters for its lifetime, the objects making up the CBDM have had to cross the cluster gas many times without much losses (unless they are destroyed and form on short time scales).\footnote{As alluded to above, and as discussed in Milgrom (2008), some degree of interaction with the gas might be present and solve the `cooling flow' problem, by the CBDM loosing kinetic energy to gas heat.}
\par
In departures from purist MOND as explanations of single clusters, the Bullet generically requires a separate explanation.
\par
What the putative CBDM is made of, how it formed, and how exactly it interacts with other cluster components are still moot questions. We have had only few constrains on it.
\par
Here I suggest that the CBDM is closely related to the recently discovered ultra-diffuse galaxies (UDGs) (Penny \& al. 2009, van Dokkum \& al. 2015a, Koda \& al. 2015), and that we may eventually draw from observations of these clues to the nature of the CBDM. It will be yet another example where two problems are better than one in the sense that the solution to the two may be the same, and that the consideration of each may help pinpoint the solution to the other.
\par
In section \ref{obs}, I describe briefly the relevant observations. Section \ref{mond} summarizes the situation in light of MOND, and section \ref{cbdm} discusses why UDGs might provide us with hints regarding CBDM.

\section{\label{obs}UDGs in the Coma cluster}
Several tens of UDGs were discovered in the Coma cluster by van Dokkum \& al. (2015a) (see also a more detailed study of one of these by van Dokkum \& al. 2015b). More recently, Koda \& al. (2015) have identified a large population ($>10^3$) of ultra-diffuse, galaxy-like objects in Coma. Even earlier,
Penny \& al. (2009) have described a small sample of similar objects in the Perseus cluster, and
Mihos \& al. (2015) have identified three such UDGs in the Virgo cluster. But, naturally, our discussion here concentrates on the much more extensive findings for Coma.
\par
These, in themselves, are rather unusual objects. As their name implies, they have very low surface brightnesses, on a par with the lowest known surface brightness disc galaxies, or with the dwarf spheroidal satellites of the Milky Way and Andromeda. However, many of them are rather larger than the previously known, typically dwarfish, low-surface-brightness galaxies.
They are not disc-like -- the distribution of their sky-projected axes ratio (whose average is 0.7-0.8) is anything but that of disc-like objects. Their surface-brightness distribution is described by a S\'{e}rsic profile with an index of 0.9-1, namely they have a nearly exponential distribution.
\par
van Dokkum \& al. (2015a) do not find UDGs within $300\kpc$ projected distance from the center of Coma (which might be due to crowding). They suggest that the UDGs are fragile and cannot survive the stronger tidal forces nearer the center. Assuming that the UDGs have been long lived in Coma, they estimate that to be immune to tidal destruction at a distance of $300\kpc$ from the center -- for which they required a tidal radius larger than twice the typical effective radius, $2r_e=6\kpc$ -- the Newtonian mass of the UDG has to be $\gtrsim 3\times 10^9\msun$, a few tens times larger than the estimated stellar mass. Koda \& al. reach similar conclusions. (Remember that the required mass scales as the cube of the radius at which robustness is required.) Interestingly, and in contrast, the UDGs that Penny \& al. (2009) could spectroscopically assign to the Perseus cluster, seem to be robust to tidal disruption, at their present position in Perseus, even without enhanced self gravity (see their fig. 7).
\par
It is not known  if the UDGs represent the extreme end of a continuum in structural properties of Coma galaxies, or are separated from `normal' galaxies in parameter space. Both van Dokkum \& al. (2015a) and Koda \& al. (2015) suggest the former, since the gap in the observed characteristics between the UDGs and `normal' Coma galaxies might be due to selection criteria. However, there is no question that in themselves, UDGs are very different, and, as discussed by both groups, must have different dynamics, and formation and evolution histories.
\par
Koda et al. (2015) deduced that the UDGs must be all but unique to clusters (Coma in the present case), since if they existed in similar proportion in the field, their number there would have to be unacceptably large.\footnote{There are very few known examples of large, low surface brightness field galaxies, such as the disc galaxies Malin 1 and NGC 7589 (see, e.g., Lelli \& al. 2010). These, however, are disc galaxies, unlike the UDGs.} Their formation and evolution must then be unique to clusters.
\par
For understanding the nature of the UDGs and their dynamics, it is important to know their full population, and whether there are many that are undergoing disruption, how many, to what extent, and how the phenomenon depends on location in the cluster (see below).
It is, thus, particularly noteworthy that van Dokkum et al. and Koda \& al. say that they left out, by intentional selection (to eliminate spurious detections), objects that show signs of tidal influences (such as tails).
This means that many babies may have been thrown out with the bathwater.
\par
There is also no information yet on the internal dynamics of the UDGs: whether they rotate, and what the internal velocities are. This would also be crucial for understanding them.
\par
The origin, formation mechanism(s), and survival of the UDGs are still a mystery. Both groups discuss various evolutionary scenarios, specific to UDGs (see Sec. \ref{infall}).

\section{\label{mond}UDGs and MOND}
In their fig 4(d), Koda \& al. (2015) plot the UDGs in the  plane of R-band-luminosity vs. $r_e$ (enclosing half of the projected light). The plotted UDGs lie roughly along a line of constant surface density.
The central line corresponds to
 \beq L_R\approx 7\times 10^7(r_e/3\kpc)^2\lsun,\eeqno{i}
and the points lie within a factor of $\sim 4$ on both sides of the line.
This distribution of values may well be an artefact of the imposed selection cutoff on the effective surface brightness of $>25 {\rm mag/arcsec}^2$, and the study's detection limit of 28 mag/arcsec$^2$. So there may be many more UDGs beyond these boundaries.
\par
For UDGs satisfying relation (\ref{i}),
the Newtonian acceleration at $2r_e$ is \footnote{This radius encloses much of the light--the light profile being roughly exponential}
 \beq\gN/\az\approx 2\times 10^{-3}(M/L), \eeqno{ii}
where $\az$ is the canonical value of $\az=1.2\times10^{-8}\cmss$, and $M/L$ is in solar units. Since this is much smaller than unity, we are dealing with deep-MOND systems. At such low accelerations, MOND predicts a mass discrepancy, which at $2r_e$ is
\beq \eta\equiv g_i/\gN\approx (\gN/\az)^{-1/2}\approx 20(M/L)^{-1/2}, \eeqno{guta}
with a factor of 2 on each side for the whole plotted population. Here, $g_i$ is the internal (MOND) acceleration in the UDG.
\par
Such large discrepancies would be predicted in standard, purist MOND, had the UDGs been isolated, and they would be broadly consistent with the requirements of tidal robustness, because with an $M/L$ of a few solar units, the internal accelerations are a few tens times larger than the estimated Newtonian acceleration. This suffices, according to the analysis of van Dokkum \& al. (2015a) and Koda \& al. (2015) (who considered extreme, not typical, cases); see Sec. \ref{obs}.
\par
However, in a cluster environment, where the UDGs are falling in the rather higher acceleration field of the cluster, the MOND external-field effect (EFE) may be acting (e.g., Milgrom 1983a, Bekenstein \& Milgrom 1984, Wu \& al. 2008).
The effect depends on where in the cluster the UDG is, but, if it does act it is expected to reduce greatly the predicted discrepancy to only between none to a few (see more details in Sec. \ref{infall} below).
\par
Below I will suggest that the UDGs are, in fact, much more baryon-massive than meets the eye at present. In this case, they will not only be more tide robust, but also rather less in the deep-MOND regime.
\par
All the above assumes purist MOND whereby, among other things, the value of  $\az$ is universal. As mentioned in the introduction, it has been suggested by different authors that the remaining discrepancy in the cores of clusters could be removed if the effective value of $\az$ there is typically about ten times larger than the canonical value (e.g., Zhao \& Famaey 2012, Khoury 2015). In the Coma, specifically, it would have to be even larger, perhaps several tens times the canonical value, to account for the dynamics in the core with only the presently identified baryons as the source of gravity.
\par
While I prefer the purist and minimalist approach, and see no compelling reasons to depart from it, I note in the present connection that with a much elevated value of $\az$, as required to remove the cluster discrepancy without additional matter, even with an EFE we would predict large mass discrepancies in UDGs.
\par
But what value of $\az$ should we use for the internal dynamics of a galaxy embedded in the external field of a cluster, the one for the galaxy or that for the cluster? The specific theory should answer this question.
Here, I note that `normal', non UDGs in clusters do not show very large mass discrepancies. But they would have, had we used a much larger value of $\az$ both internally and externally. Thus, if one chooses this rout to solve the cluster conundrum, one will have to say that the canonical $\az$ should be used internally when the `cluster' value is used externally. (This does not seem to be the case by the suggestion of Zhao \& famaey 2012, where the value $\az$ depends on the local value of the gravitational potential, since the galactic contribution to the potential is rather smaller than that of the cluster, so the `cluster $\az$' should also apply inside cluster galaxies.)

The other concern with this approach is that in the outskirts of Coma only a rather smaller value of $\az$ is needed, perhaps three times the canonical value. On the other hand, the requirements on tidal robustness are more lax at larger cluster radii.
\par
Be the case as it may, below, I consider the implications of `minimalist' MOND, which does not fully explain away the mass discrepancies in UDGs, and which does require some sort of CBDM.
\subsection{\label{infall}Are UDGs now falling into Coma for the first time?}
Consider now the possibility that the UDGs are not fully immune to tidal destruction, but are only now infalling into Coma for the first time, and so have not yet had time to disintegrate. (Tidal disruption times are comparable to the free-fall time in the cluster.)
van Dokkum \& al. (2015a) discuss this possibility, and give it a low weight, based on Newtonian considerations. They say: `It is not clear how UDGs were formed. It seems unlikely
that they are the product of galaxy harassment...or tidal stirring...of infalling galaxies: these  processes tend  to shrink galaxies,  as  the  stars  at larger radii are less bound than the stars at small radii...We note, however, that the morphological evolution of infalling galaxies is difficult to predict, as it probably
depends sensitively on the shape of the inner dark matter profile...'
\par
However, in MOND the situation is rather different. There are processes at work, totally absent from standard dynamics, that could greatly affect the appearance of infalling low-surface-density galaxies. Such a notable MOND effect is the above mentioned EFE. This effect may cause galaxies to inflate appreciably as they fall into Coma, and so counter and perhaps outstrip possible shrinking due to loss of the outer-lying stars due to tidal effects.\footnote{It would be instructive to learn if and how the UDG size correlates with distance from the Coma center (after removal of obvious selection biases).}
\par
The exact action of the MOND EFE depends on the specific MOND theory used (examples, e.g., in Milgrom 1983a, Bekenstein \& Milgrom 1984, Milgrom 2010). Consider the case where the acceleration, $\vg_e$, with which the subsystem is falling, is much larger than its internal accelerations, $\vg_i$ ($g_i \equiv|\vg_i|$)--as happens here for a wide range of the parameter. The generic effect implies that the internal dynamics is quasi-Newtonian, only with an effective gravitational constant
\beq G_e\approx G/\m(g_e/\az), \eeqno{juop}
and also some anisotropy of order unity that is introduced by the direction of the external field. Here, $\m(x)$ is the MOND interpolating function, and $g_e\equiv|\vg_e|$. In the very inner parts of Coma, say within $100\kpc$ of the center, $g_e\gg\az$, and so $\m\approx 1$, and the internal dynamics is Newtonian. But for larger radii, $\m<1$, so $G_e>G$, and, furthermore, it is position dependent.
\par
The general problem of solving, in MOND, the evolution of a UDG as it falls into a cluster, including the effects of tidal forces, without and with various amounts of CBDM, and for different initial conditions and orbits, can be treated straightforwardly with a MOND N-body code, and will make an interesting project.
\par
Brada \& Milgrom (2000) have discussed the generalities of what happens to the internal structure of a small daughter system (there a dwarf satellite, here a UDG) that is falling in the field of a mother system (there a galaxy, here a cluster), as a result of the variations of $\vg_e$ along the orbit. They also considered several numerical examples.
\par
Here I consider the problem only in semi-quantitative terms, and make the following assumptions:
(i) I ignore the secondary, anisotropy effect, and consider, approximately, only the dominant effect of the variation in $g_e$, and hence of  $G_e$, during infall. (ii) Assume that $g_e\gg g_i$, so the effect can, indeed, be summarized by a position-dependent renormalization of $G$. This was not necessarily the case for an infalling UDG during the earlier stages of its infall, where $g_e$ was smaller, and $g_i$ larger (see why below). But it is valid for the UDGs reported by van Dokkum \& al. and Koda \& al. at their present state. (iii) The effects of varying $G_e$ may be considered adiabatic, namely that the free-fall time in the cluster is longer than intrinsic times, such as the orbital time of a test particle in the system. This approximation was better in the earlier stages of the infall, but becomes worse and breaks down at later stages (see estimates below for the reported UDGs). (iv) The system does not exchange mass with its surrounding (e.g., no tidal losses), so its mass is fixed (v) The system is at all times in virial equilibrium.
\par
It is not easy to satisfy all these requirements; so, indeed, numerical studies are necessary. My only purpose here is to explain, in isolation, the inflation of an infalling system, resulting from the EFE
\par
If has then been shown in Brada \& Milgrom (2000) that if one makes all the above assumptions, the subsystem adjusts it size, $r$, and internal velocities, $v$, to the value of the external field according to
\beq r\propto v^{-1}\propto G_e^{-1}\propto \m(g_e/\az). \eeqno{maty}
Milgrom 2015 (see especially footnote 10 there), treated the equivalent problem analytically in more detail, with the same conclusions, showing, furthermore, the size variation is homologous, and the velocity variation is uniform across the subsystem.
\par
With these scalings, the internal accelerations during infall decrease, and scale
as $v^2/r\propto G_e^3$, the internal dynamical times increase, and scale as
$r/v\propto G_e^{-2}$. So we see that the initial internal accelerations may have been rather larger, and the dynamical time scales rather shorter, than they are after some infall.
The inflation of the system during infall makes it even more vulnerable to tidal disruption than it would have been, starting from the same initial conditions in Newtonian dynamics (Brada \& Milgrom 2000).
\par
For the Coma acceleration field, we can use, for example, the estimates from  fig. 8 of Lokas \& Mamon (2003) of dynamical masses, from galaxy kinematics (other studies would give similar estimates, see below). One finds within $R=2\mpc$ of the center, a dynamical mass of $M_d\approx 10^{15}\msun$.
The measured acceleration at this radius is thus
\beq g_e\approx\frac{M_dG}{R^2}=3.5\times 10^{-9}M_{15}\approx 0.3M_{15}\az \eeqno{iii}
(where, $M_{15}=M_d/10^{15}\msun$).
With the oft used interpolating function, $\m(x)=x/(1+x)$, which, for $x\lesssim 20$, relevant here, has proved successful in accounting for rotation curves (e.g., Famaey \& McGaugh 2012) and dynamics of ellipticals (Milgrom 2012)\footnote{There are other interpolating functions that give practically identical results for $x<20$, but which approach unity for $x\rar\infty$ much faster, as required by Solar-System constraints (see e.g., Milgrom 2012).}--giving $\m(g/\az)\approx 0.22$.
At $1\mpc$ of Coma's center, the same figure gives $M_d\approx 6\times 10^{14}\msun$, which corresponds to $g/\az\approx 0.7$ or $\m(g/\az)\approx 0.4$, and
at $0.3\mpc$, $M_d\approx 1.7\times 10^{14}\msun$, which corresponds to $g/\az\approx 2.2$, or $\m(g/\az)\approx 0.7$. And, $\m(g/\az)$ approaches 1 deeper in.
\par
Thus, as long the underlying assumptions for eq.(\ref{maty}) are met, a system's size could increase in falling from $2\mpc$ by a factor of 2 at $1\mpc$, a factor of 3.5 at $0.3\mpc$, and a factor of $\sim 5$ further down; and this on top of the increase suffered during infall to $R=2\mpc$.
This process of inflation started roughly at the position where the internal accelerations dropped below the external ones: $g_i\sim g_e$.  By eq. (\ref{maty}) we have within the external-field-dominance region
\beq g_i/g_e\propto G_e^3/g_e\propto g_e^{-1}\m^{-3}(g_e/\az).  \eeqno{bana}
Beyond $2\mpc$, we are already in the MOND regime, where $\m(x)\sim x$, and, furthermore, the (MOND) acceleration decreases as $R^{-1}$, since most of the Coma (baryonic) mass is within $2\mpc$. We then have there $g_i/g_e\propto R^4$. Thus, even if at $2\mpc$, $g_i$ (which is difficult to estimate) is an order of magnitude smaller than $g_e$ (see example below), inflation is at work only below $\sim 4\mpc$.

\par
To assess adiabaticity we may roughly estimate the free-fall time say from several $\mpc$ to $1\mpc$ as $\tau\sim 1\mpc/1500\kms=2\times10^{16}\sec\approx 6\times 10^8 {\rm y}$. If UDGs are destroyed typically over such a time, we see that $\gtrsim 10^4$ UDGs may have unloaded their cargo onto Coma over the Hubble time.
\par
It is difficult, at present, to estimate internal properties of the UDGs (orbital times and accelerations), since we do not have any direct knowledge of their internal dynamics. We do not know what the internal velocities are, whether UDGs contain matter in addition to the observed baryons, and whether the UDGs are now in virial equilibrium.
\par
To get a sense of the numbers, let us estimate the internal (MOND) acceleration at $2r_e$ in a UDG sitting on relation (\ref{i}) and located $2\mpc$ from the Coma center. Suppose that its real (baryonic) mass is $q$ times the observed stellar mass (due to the presence of CBDM). Then, instead of the expression in eq.(\ref{ii}), its Newtonian internal acceleration at $2r_e$ is $\gN/\az\approx 2\times 10^{-3}q(M/L)$, while the MOND acceleration is enhanced by a factor $1/\m(g_e/\az)\approx 5$; so $g_i/\az=g^i_M/\az\approx  10^{-2}q(M/L)$, while $g_e/\az\approx 0.3$ there. So, $g_i/g_e\approx  q(M/L)/30$. Such a UDG is external-field dominated for $q(M/L)\lesssim 30$. In this case, it would have been so if moved adiabatically up to a Coma radius of $\sim 2  [q(M/L)/30]^{-1/4}\mpc$ [from eq.(\ref{bana}) and the argument that follows it].
\par
To assess adiabaticity we would need the measured internal velocities, which we do not have. If we assume that the UDG is in virial equilibrium, we can estimate its internal virial speed from the Newtonian virial relation with the effective, MONDian, gravitational constant, $G_e$. This gives $v_i\sim (g_ir)^{1/2}$, where $g_i$ is given above. From this, the orbital time, also at $2r_e$, can be estimated as
$\tau\equiv 2\pi r/v_i$. For the above example, and taking $r_e=3\kpc$, we have $\tau\sim  8\times 10^{16}q^{1/2}(M/L)^{-1/2}\sec$.
This is to be compared with the free-fall time, e.g., in our estimate above.
We see that the two are comparable for $q(M/L)\sim 10$.
This stresses all the more that our assumptions above are at best marginal, and that a numerical study is required.
\par
One purpose of such studies would be to find the initial state of the UDG progenitors and see whether we can identify them in the vicinity of present-day clusters, or in the field.
To do this without the trial and error afforded by simulations would be tantamount to running the UDG evolution back in time, starting from their present position and internal state as initial conditions.
This, however, is not possible at present, for several reasons (see above assumptions):
(i) This evolution involves irreversible processes, and can thus not be time reversed. Two contributors to this are: a. mass losses due to , e.g., tidal effect, and b. the fact that the changes of the external field, and hence the EFE, are not adiabatic.
(ii) In addition, we do not really know what the present state of these systems is: There is no information on the intrinsic velocities, we do not even know if the UDG are in virial equilibrium, and  we do not know how much CBDM they contain.
Instead, simulations can start from disparate initial states of UDG progenitors, with various structures and CBDM contents, and let these fall into a Coma-like environment, and see if their present appearance, as far as we can characterize it, matches the observed ones.
\par
There exist already efficient numerical codes based on two versions of MOND as modified gravity, that can tackle these problems (Candlish, Smith \& Fellhauer 2015; L\"{u}ghausen, Famaey \& Kroupa 2015).
\par
As already mentioned, it is difficult to know what fraction of UFGs and their kin show signs of tidal effects, since both discovery groups have a priori excluded from their selected candidates those that show such signs. This would be very important to know.
Mihos \& al. (2015) report that one of their three ultra-diffuse objects in Virgo `appears to be ... in the process of being tidally stripped.'

\section{\label{cbdm}UDGs, key to CBDM?}
If individual UDGs are long lived in Coma and other galaxy clusters, they could indeed point and conduce to the existence of CBDM as the solution of the MOND cluster conundrum.
\par
The mass discrepancy in Coma that remains in MOND may be estimated, say within $2\mpc$ of the center, using the results of Lokas \& Mamon (2003) given above: $M_d(R<2\mpc)\approx 10^{15}\msun$. Another estimate of this mass is by Gavazzi et al. (2009; their table 1), from weak lensing: Within their $R=r_{200}\sim 2\mpc$, they give the range $M_d\approx(0.5-1.5)\times 10^{15}\msun$. In MOND, the required (baryonic) mass to produce such a Newtonian dynamical mass is
\beq M_M(R<2\mpc)\approx M_d(R<2\mpc)\m[g_e(2\mpc)/\az], \eeqno{jasyu}
with $g_e$ from eq.(\ref{iii}). This gives [for $M_d(R<2\mpc)= 10^{15}\msun$] $ M_M(R<2\mpc)\approx 2.2\times 10^{14}\msun$.
The contribution of the presently known baryons (hot gas plus stars) within this radius is estimated in Lokas \& Mamon (2003) at $M_b\approx 1.4\times 10^{14}\msun$. We see then that the remaining discrepancy at this radius is indeed, roughly, a factor of 2, and MOND still requires some $\sim 10^{14}\msun$ of CBDM within $2\mpc$, compared with the DM mass of $\sim 9\times10^{14}\msun$ that is required in Newtonian dynamics.\footnote{Note that at smaller radii the relative MOND discrepancy is rather higher. See e.g. fig. 1 of Angus, Famaey, and Boute (2008).}
\par
Very similar numbers come out in the MOND analysis of Angus, Famaey, \& Boute 2008 (see their fig. 1), for the rich and massive cluster A2029. For their largest plotted radius of $\approx 1.5\mpc$ one reads
$M_d\sim 1\times 10^{15}\msun$, from which one deduces, as above, $M\_M\sim 3\times 10^{14}\msun$, compared with their baryonic mass at this radius, estimated at $M_b\sim 1.5\times 10^{14}\msun$, requiring roughly this same mass in CBDM.\footnote{Angus \& al. (2008) also analyzed another, rather less massive cluster, A2717, within $800\kpc$ where they find a Newtonian, dynamical mass of $\sim 1\times 10^{14}\msun$, a MOND dynamical mass of $\sim 2.5\times 10^{13}\msun$, baryonic mass of $\sim 1.5\times 10^{13}\msun$, leaving a required CBDM mass of $\sim 1\times 10^{13}\msun$. Similar proportions of masses, with rather lower absolute values are found for x-ray groups.}
Even earlier, Sanders (1999) has analyzed a large sample of x-ray clusters in MOND, and found similar MOND discrepancies (see his fig. 2).
\par
It is instructive to recall, here, a lesson from the history of MOND that might repeat itself if the surmise of CBDM is correct. The (global) MOND analysis of clusters at the very advent of MOND (Milgrom 1983b), gave rather large MOND mass-to-light ratios of several of the clusters studied (a few tens in solar units), among them Coma and the above mentioned A2029. At the time it was thought that galaxies (stars) alone make up the baryonic mass in clusters. So these large $M/L$ values, as compared with a value of a few for stars, would have meant remaining MOND mass discrepancies of order ten.
It was suggested at the time, based on the fact that clusters were known to be x-ray sources, that `If... we allow for intergalactic gas (as evidenced by x-ray emission), the results are consistent with most of the dynamic mass in clusters being of conventional forms.' Indeed, a few years later, with the advent of the Einstein x-ray satellite, it was discovered (Abramopoulos \& Ku 1983, Jones \& Forman 1984) that the x-ray emitting gas contributes much more than stars to the baryonic mass, vindicating the above prediction, if not fully, by reducing the remaining MOND discrepancy from $\sim 10$ to $\sim 2$.
The surmised CBDM is to bridge this remaining factor of 2, and are still waiting to be discovered.
\par
While possible, it is a stretch to assume that all the $\sim 10^{14}\msun$ of CBDM resides in the $\sim 10^3$ UDGs reported by Koda et al., as this would require them to have an average mass of $10^{11}\msun$. As mentioned in Sec. \ref{obs}, there may be, at present, many more UDGs in Coma than reported by Koda et al., considering the selection cuts they applied, and the detection limits. So the mass of each could be rather smaller.
\par
It seems, however, more likely that much of the CBDM is now dispersed in the cluster at large, perhaps having been formed in the UDGs or their progenitors.
\par
We saw in Sec. \ref{infall}, that if, for example, UDGs are infalling and are being disintegrated within about one Coma free-fall time, then over the Hubble time some $\gtrsim 10^4$ such objects may have assimilated into Coma's intracluster medium. This would require them to have had, on average, only $\lesssim 10^{10}\msun$ to account for the required CBDM mass. Whether with such masses they are vulnerable to tidal disruption depends on their initial size.
MOND, as we saw, and as discussed in Brada \& Milgrom (2000), makes them more so.
\par
We will be able to understand these aspects much better with the help of detailed numerical calculations, and if the internal dynamical state of the UDGs (e.g., their internal velocities) are measured.
\par
It is also important to ascertain that UDGs are as common in other rich clusters as in Coma.
\par
Beyond the above important, quantitative questions that remain open, the discovery of many UDGs in Coma lends more plausibility to the CBDM solution of the MOND cluster conundrum:
In the first place it is another indication that we have not fully understood the baryonic content of x-ray clusters, and how its different components form, are destroyed, and exchange mass with each other.
Secondly, if before we had had to suppose that the yet-undetected constituents of the CBDM (stellar remnants? cold gas clouds?) formed in the intracluster gas, UDGs offer a rather more plausible formation sites. The `normal' galaxies within clusters could also be the formation site, but they are not unique to clusters, whereas UDGs, like the CBDM, are. Also, the former are not suspected to be heavy with `DM', while the latter are.
\par
If the cluster DM is indeed of the CBDM envisaged here, it would also be easy to understand, why, in the Bullet Cluster, the CBDM followed the galaxies in the aftermath of the collision, and has not stuck with the gas in the collision zone.
\par
Thus the UDGs help direct our thoughts and focus on more specific pathways to understanding what the CBDM could be.

\end{document}